\documentclass[%
 aip,
 jcp,
 amsmath,amssymb,
reprint,%
]{revtex4-1}

\usepackage{graphicx}
\usepackage{dcolumn}
\usepackage{bm}

\usepackage[utf8]{inputenc}
\usepackage[T1]{fontenc}
\usepackage{mathptmx}
\usepackage{etoolbox}

\usepackage{caption}
\usepackage{subcaption}
\usepackage{xcolor}
\newcommand{\diff}{\mathrm{d}}

\newcommand{\marker}[1]{\parbox[c][5.5ex][c]{4ex}{\vspace{0ex}\includegraphics[height=4ex]{#1}\vspace{0ex}}}
\usepackage{multirow}
\usepackage{comment}

\makeatletter
\def\@email#1#2{%
 \endgroup
 \patchcmd{\titleblock@produce}
  {\frontmatter@RRAPformat}
  {\frontmatter@RRAPformat{\produce@RRAP{*#1\href{mailto:#2}{#2}}}\frontmatter@RRAPformat}
  {}{}
}%
\makeatother
\begin{document}

\preprint{AIP/123-QED}

\title{Vacancy defects in square-triangle tilings and their implications for quasicrystals formed by  square-shoulder particles}

\author{Alptu\u g Ulug\" ol}
 \affiliation{%
Soft Condensed Matter and Biophysics Group, Debye Institute for Nanomaterials Science, Utrecht University, Princetonplein 1, Utrecht, 3584 CC, Netherlands
}%
\affiliation{%
Authors contributed equally.
}
\email{a.ulugol@uu.nl}

\author{Giovanni Del Monte}
 \affiliation{%
Soft Condensed Matter and Biophysics Group, Debye Institute for Nanomaterials Science, Utrecht University, Princetonplein 1, Utrecht, 3584 CC, Netherlands
}%
\affiliation{%
Authors contributed equally.
}

\author{Eline K. Kempkes}%
\affiliation{%
Soft Condensed Matter and Biophysics Group, Debye Institute for Nanomaterials Science, Utrecht University, Princetonplein 1, Utrecht, 3584 CC, Netherlands
}
\affiliation{Computational Soft Matter Lab, Computational Chemistry Group and Computational Science Lab, Van ’t Hoff Institute for Molecular Sciences and Informatics Institute, University of Amsterdam, Science Park 904, 1098 XH Amsterdam, The Netherlands.}

\author{Frank Smallenburg}%
\affiliation{Laboratoire de Physique des Solides, Université Paris-Saclay, Orsay, 91405, France}

\author{Laura Filion}
\affiliation{%
Soft Condensed Matter and Biophysics Group, Debye Institute for Nanomaterials Science, Utrecht University, Princetonplein 1, Utrecht, 3584 CC, Netherlands
}%

\date{\today}

\begin{abstract}
Almost all observed square-triangle quasicrystals in soft-matter systems contain a large number of point-like defects, yet the role these defects play in stabilizing the quasicrystal phase remains poorly understood.  In this work, we investigate the thermodynamic role of such defects in the widely observed 12-fold symmetric square-triangle quasicrystal. We develop a new Monte Carlo simulation to compute the configurational entropy of square-triangle tilings augmented to contain two types of irregular hexagons as defect tiles. We find that the introduction of defects leads to a notable entropy gain, with each defect contributing considerably more than a conventional vacancy in a periodic crystal. Intriguingly, the entropy gain is not simply due to individual defect types but is amplified by their combinatorial mixing. We then apply our findings to a microscopic model of core-corona particles interacting via a square-shoulder potential. By combining the configurational entropy  with vibrational free-energy calculations, we predict the equilibrium defect concentration and confirm that the quasicrystalline phase contains a higher concentration of point-defects than a typical periodic crystal. These results provide a new understanding of the prominence of observed defects in soft-matter quasicrystals.
\end{abstract}

\maketitle

\section{Introduction}

Quasicrystals (QCs) are unusual states of matter characterized by long-range order while lacking the translational symmetry of conventional crystals. At the same time, they can exhibit rotational symmetries that are forbidden in periodic lattices.
The first experimental evidence for quasicrystalline phases was discovered in metallic alloys \cite{shechtman1984metallic}. Since then, QCs have been observed in a wide range of soft matter systems \cite{fayen2023self, zeng2004supramolecular, zeng2023columnar, forster2013quasicrystalline,forster2020quasicrystals,hayashida2007polymeric,li2023growth,liu2019rational,urgel2016quasicrystallinity,schenk20222d,plati2024quasi}, including dendrimers~\cite{zeng2004supramolecular}, star polymers~\cite{hayashida2007polymeric}, block copolymers~\cite{gillard2016dodecagonal}, and nanoparticle assemblies~\cite{talapin2009quasicrystalline,ye2017quasicrystalline}.
Alongside these experimental advances, several minimal interaction models have been proposed to reproduce quasicrystalline order in numerical simulations. These include systems governed by isotropic pair potentials \cite{dotera2014mosaic,engel2007self,kryuchkov2018complex,padilla2020phase}, patchy particles \cite{noya2025one,reinhardt2013computing,tracey2021programming,van2012formation,reinhardt2016self,noya2021design,gemeinhardt2019stabilizing}, and multicomponent mixtures \cite{scacchi2020quasicrystal,fayen2024quasicrystal,fayen2020infinite,fayen2023self,bedolla2024inverse}. Among these, one of the most widely studied is the square-shoulder potential, which describes core-corona particles interacting via a short-range hard repulsion surrounded by a softer, finite-range repulsive shoulder. This model has been shown to form a wide range of quasicrystals characterized by different symmetries, including e.g. 10, 12, 18, and 24-fold rotational order \cite{dotera2014mosaic}.

The structure of quasicrystals can typically be represented by tilings of space using a finite set of regular polygons that appear at specific orientations and share a common particle decoration, for instance with particles located at the polygon vertices. Multiple distinct arrangements of these tiles are possible, which greatly enhances the thermodynamic stability of quasicrystalline phases through a positive configurational entropy contribution.
Arguably, the most common soft matter QC is a two-dimensional structure with 12-fold symmetry that is made up from random-packings of square and equilateral triangle tiles \cite{oxborrow1993random,imperor2021square,imperor2024higher,fayen2023self}. Intriguingly, such tilings rarely appear perfectly regular. Instead, essentially all observed square-triangle quasicrystals contain a large number of point-like defects, typically in the form of additional hexagonal tiles \cite{ye2017quasicrystalline,rochal2016soft,liu2022expanding,iacovella2011self,dotera2014mosaic,
ryltsev2017universal, schenk20222d, liu2019rational, van2012formation,tracey2021programming,ishimasa1988electron,dzugutov1993formation,schenk2019full,kryuchkov2018complex,talapin2009quasicrystalline,hayashida2007polymeric}.
However, it is not clear whether the high concentration of these tiles stems from kinetic effects during self-assembly, or whether they are an inherent part of the quasicrystal phase which plays a key role in their stability. 

Here, we address this question in detail for the simple case of a square-shoulder model in a parameter regime where a square-triangle quasicrystal is stable. Allowing explicitly for
the presence of hexagonal defect tiles, we investigate how changes in the number of accessible tilings and in their vibrational properties affect the thermodynamic stability of the quasicrystalline phase. This study is divided into two parts, each employing a distinct model. In the first part, we analyze the configurational entropy of defected square-triangle tilings using a variation of the tile-based Monte Carlo approach we introduced in Ref. \cite{ulugol2025defects}. In the second part, we incorporate these insights into a microscopic system of core-corona particles governed by square-shoulder interactions. By combining vibrational free-energy calculations with the configurational entropy obtained from the tile model, we determine the equilibrium defect concentrations and assess their impact on the stability of the quasicrystalline phase.

Our findings demonstrate that defect tiles, rather than disrupting the quasicrystalline order, can significantly increase the number of accessible configurations and thereby promote thermodynamic stability, leading to high equilibrium defect concentrations.

\begin{figure*}
    \centering
\includegraphics[width=\linewidth]{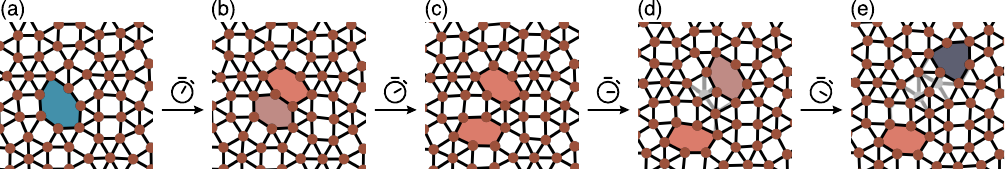}
    \caption{Time series of defect dynamics from simulations of particles interacting via a square-shoulder potential: (a) a vacancy is introduced by removing a particle; (b) the vacancy splits into a pair of left- and right-handed egg defects; (c) the lower egg propagates and changes chirality; (d) the upper egg also propagates and flips chirality; (e) the upper egg further propagates and transforms into a shield defect. In panels (d) and (e), gray edges and particles highlight differences from the original configuration shown in (a).}
    \label{fig:defectdyn}
\end{figure*}

\begin{figure}
    \centering    \includegraphics[width=\linewidth]{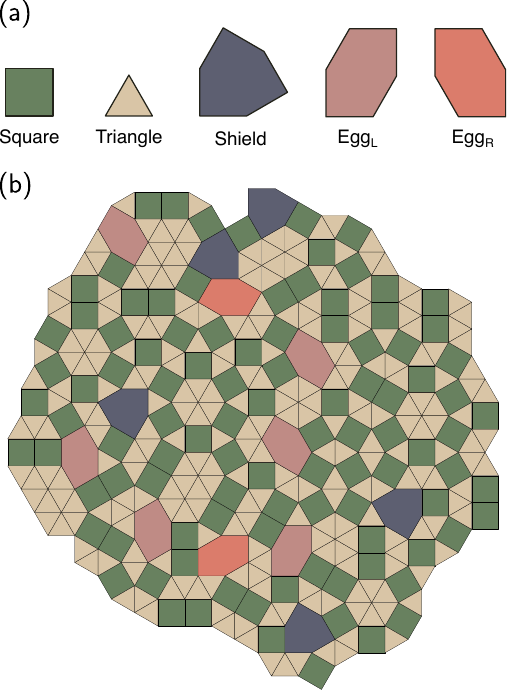}
    \caption{Schematic of the simulation geometry. (a) The five tile types allowed in the tile-based simulations: square and triangle tiles correspond to the regular tiles of the ideal quasicrystalline tiling, while the remaining three represent the only permitted defect tiles. (b) Representative snapshot from a tile-based simulation containing 289 vertices.}
    \label{fig:tiles-tiling}
\end{figure}

\section{Configurational entropy of defects}

\subsection{Model and methods}

\subsubsection{Vacancy-related defects in square-triangle tilings}

The hexagonal defect tiles we consider in this work arise naturally when a vertex is removed from a perfect square-triangle tiling. This can be seen directly in a simulation of e.g. a square-shoulder model potential, where one particle (corresponding to a vertex in the tiling) is omitted from a quasicrystalline starting configuration. As illustrated in Fig.~\ref{fig:defectdyn}, this removal leads to rearrangements of the local neighborhood, resulting in a pair of two new types of hexagonal tiles, which we refer to here as shields and eggs. Note that the egg-shaped tiles are chiral, leading to three distinct defect tile types in total: Shield, Egg$_L$, and Egg$_R$ (see Fig. \ref{fig:tiles-tiling}a).

\subsubsection{Lattice-based simulation model}

To study the configurational entropy of square-triangle tilings with shield or egg-shaped defects, we adapt the methodology we introduced in Ref.~\cite{ulugol2025defects}. Specifically, we use an idealized tile-based model, where the tiling consists exclusively of square, triangle, shield, and egg tiles, as shown in Fig.~\ref{fig:tiles-tiling}(a). To control the number of defects in the system, we introduce a finite free-energy penalty $\epsilon_D$ per defect tile.  Note that we have assigned all defect types to have the same energy cost. Additionally, we impose open-boundary conditions by incorporating a line tension $\gamma$, which is proportional to the perimeter $L$ of the tiling. The energy $U_t$ of a specific tiling configuration $t$ is then given by:
\begin{equation}
    U_t = n_D\epsilon_D + \gamma L,
\end{equation}
where $n_D$ is the total number of shield and egg tiles. Throughout the fist part of this study, we work in the $N\gamma T$ ensemble, where the number of vertices ($N$), line tension ($\gamma$), and temperature ($T$) are fixed.

\subsubsection{Monte Carlo simulation}

The Monte Carlo (MC) algorithm we use is designed to sample configurational space while ensuring that the tiling contains no holes or overlaps. 
In order to design a Monte Carlo move that samples the desired ensemble, we first consider the fact that only a small fraction of vertices in the system can move (most are locked in place by their local environment). Hence, to efficiently identify possible moves, we first identify so-called \emph{mobile edges}, which can move to allow for local rearrangements by rotating around one of the two vertices it connects. If an edge is in the bulk and has only squares and triangles on its sides, any rotation of the edge results in an invalid tiling due to either overlaps or invalid tiles rendering the edge immobile. Therefore, mobile edges are edges that are either on the boundary or part of a shield/egg tile. To perform the edge rotation MC move, we 
\begin{enumerate}
    \item choose a random mobile edge,
    \item choose one of the ends randomly as the pivot point,
    \item choose a random direction, i.e., clockwise or counter-clockwise,
    \item rotate the edge around the chosen pivot point by $\pi/6$ in the chosen direction. Note that this displaces the vertex on the other side of the edge.
\end{enumerate}
If the resulting tiling is invalid, we reject the move. If it is valid, we accept the move with the acceptance probability,
\begin{equation}
    P_\mathrm{acc}(i\to f)=
        \min\left(1, \frac{\ell_i}{\ell_f}e^{-\beta\Delta U}\right), 
\end{equation}
where $i,f$ denote the initial and final configuration, respectively. Additionally, $\ell_i$ and $\ell_f$ denote the number of mobile edges in the configurations, $\Delta U = U_f - U_i$ denotes the potential energy difference between the configurations, and  $\beta = 1/k_BT$ with $k_B$ Boltzmann's constant. This move alters the local tiling configuration without changing the number of vertices while allowing tile type changes.

Using this approach, we simulate square-triangle-shield-egg tilings with edge length $a$. We vary the defect penalty $\beta\epsilon_D \in [-1,7]$, line tensions $\beta\gamma a \in \{3.5,4,4.5,5,5.5\}$, and  the number of vertices $N\in \{2121,4123,8004,14068\}$. 
Each simulation is initialized with a randomized $[D,I]$-Stampfli inflationary tiling \cite{stampfli1986dodecagonal}, which provides a well-defined square-triangle tiling to start with. The system then evolves through Monte Carlo updates.
Once equilibrium is reached, we start collecting statistics of tile counts and orientations, edge orientations, and the length of the perimeter.  

This simulation framework provides a direct way to quantify how defects influence the configurational entropy of square-triangle tilings, enabling the analysis of defect energetics and equilibrium concentrations.

\subsubsection{Configurational entropy of the tiling}

A central question in our study is how the presence of shield and egg defects affects the total configurational entropy of the system. Specifically, we seek to quantify the configurational entropy difference between square-triangle tilings with defects and a defect-free square-triangle tiling. We define this entropy difference as:
\begin{equation}
\label{eq:deltaS}
    \Delta S(N, \gamma, T, \epsilon_D) = S(N, \gamma, T, \epsilon_D) - S_{\mathrm{sq-tr}}(N, \gamma, T),
\end{equation}
where $S(N, \gamma, T, \epsilon_D)$ is the configurational entropy of a system containing a finite concentration of defects (regulated by the defect energy $\epsilon_D$), and $S_{\mathrm{sq-tr}}$ corresponds to the configurational entropy of defect-free square-triangle tilings, which is recovered in the limit $\epsilon_D/k_B T \to \infty$. In the thermodynamic limit ($N \to \infty$), boundary effects become negligible, allowing us to express the configurational-entropy difference as:
\begin{equation}
    \Delta S = N \Delta s(\beta \epsilon_D),
\end{equation}
where $\Delta s$ is the entropy difference per vertex between a pure square-triangle tiling and a square-triangle-shield-egg tiling.

To compute $\Delta S$, we first determine the free-energy difference $\Delta F$, which is related to the entropy via:
\begin{equation}\label{eq:F=U-TS}
    \Delta F = \Delta U - T \Delta S.
\end{equation}
Here, $\Delta U$ represents the potential energy difference between a system with finite $\epsilon_D$ and a defect-free tiling, and is given by:
\begin{eqnarray}\label{eq:DeltaU}
    \Delta U(N, \gamma, T, \epsilon_D) &=& 
    \epsilon_D \left\langle n_D \right\rangle_{N,\gamma,T,\epsilon_D} + \nonumber\\
    && \gamma \left(
    \left\langle L \right\rangle_{N,\gamma,T,\epsilon_D} -
    \left\langle L \right\rangle_{N,\gamma,T,\infty}\right),
\end{eqnarray}
where $n_D$ denotes the total number of shield and egg defects in the system. Using thermodynamic integration, $\Delta F$ can be computed as:
\begin{equation}\label{eq:thermo-integrate}
\begin{aligned}
   \Delta F(N,\gamma,T,\epsilon_D) 
   &= -\int_{\epsilon_D}^\infty\diff\epsilon_D^\prime\left\langle \frac{\partial U}{\partial\epsilon_D} \right\rangle_{N,\gamma,T,\epsilon_D^\prime},\\
    &= -\int_{\epsilon_D}^\infty\diff\epsilon_D^\prime\left\langle n_D\right\rangle_{N,\gamma,T,\epsilon_D^\prime}.
\end{aligned}
\end{equation}
Here, $\langle\ldots\rangle$ denotes an ensemble average over equilibrium simulations.

To obtain the total configurational entropy of the tiling (i.e. $S(N, \gamma, T, \epsilon_D)$ in Eq.~(\ref{eq:deltaS})), we combine $\Delta S$ with the known configurational entropy of the defect-free square-triangle tiling. In the thermodynamic limit, this configurational entropy has been determined via the Bethe ansatz \cite{widom1993bethe,kalugin1994square}, yielding:
\begin{equation}\label{eq:sconf-def-free}
    \frac{S_{\mathrm{sq-tr}}}{Nk_B} = \log(108) - 2\sqrt{3}\log\left(2+\sqrt{3}\right) =  0.12005524 \ldots 
\end{equation}

\subsubsection{Configurational entropy at low defect concentration}
To obtain a better insight into the effect of defects on entropy, we consider the regime where shield and egg defects are rare and do not interact. In this low-concentration limit, the total configurational entropy can be approximated by analyzing the leading-order terms in the partition function. We begin by expressing the partition function in the large-system-size limit, where boundary contributions are negligible:
\begin{equation}
    Z(N,T,\epsilon_s,\epsilon_e) = \sum_{n_s,n_e} \Omega_{(n_s,n_e)} \, e^{-\beta(\epsilon_s n_s + \epsilon_e n_e)},
\end{equation}
where $n_s$ and $n_e$ are the number of shields and eggs, respectively; $\epsilon_s$ and $\epsilon_e$ are their corresponding energy penalties; and $\Omega_{(n_s,n_e)}$ denotes the number of distinct configurations with $n_s$ shield and $n_e$ egg tiles given $N$ vertices.  Note that here we are treating the shields and eggs as separate defect types with their own energy penalty (in contrast to what we have done until now). Additionally, we group both chiralities of eggs together such that $n_e$ is the total number of egg defects.

In the limit of large defect cost, the partition function can be expanded to first order in $e^{-\beta\epsilon_{s(e)}}$:
\begin{equation}
\begin{aligned}
    Z(N,T,\epsilon_s,\epsilon_e) 
    &\approx \Omega_{(0,0)} + \Omega_{(1,0)} e^{-\beta\epsilon_s} + \Omega_{(0,1)} e^{-\beta\epsilon_e} \\
    &= e^{S_{\mathrm{sq\text{-}tr}}/k_B} + e^{-\beta\epsilon_s + S_{(1,0)}/k_B} + e^{-\beta\epsilon_e + S_{(0,1)}/k_B} \\
    &= e^{S_{\mathrm{sq\text{-}tr}}/k_B} \left(1 + \langle n_s \rangle + \langle n_e \rangle \right),
    \label{eq:Znt}
\end{aligned}
\end{equation}
where $S_{\mathrm{sq\text{-}tr}} = k_B \log \Omega_{(0,0)}$ is the entropy of the pure square-triangle tiling, and
\begin{equation}
    S_{(n_s,n_e)} = k_B \log \Omega_{(n_s,n_e)}
\end{equation}
is the configurational entropy of a tiling containing $n_s$ shields and $n_e$ eggs.

In this approximation, the equilibrium defect concentrations follow from minimizing the free energy, yielding:
\begin{equation}\label{eq:defect-scaling-high-epsilon}
    \langle n_s \rangle = e^{-\beta \epsilon_s + \Delta S_s / k_B}, \quad 
    \langle n_e \rangle = e^{-\beta \epsilon_e + \Delta S_e / k_B},
\end{equation}
with
\begin{equation}
    \Delta S_s := S_{(1,0)} - S_{\mathrm{sq\text{-}tr}}, \quad
    \Delta S_e := S_{(0,1)} - S_{\mathrm{sq\text{-}tr}}
\end{equation}
representing the entropy gained from adding a single shield or egg tile, respectively.  Equation \ref{eq:defect-scaling-high-epsilon} allows us to extract entropy gains from simulations by measuring the equilibrium defect concentration at high defect energy costs.

Since the number of defects scales extensively, $\Delta S_{s(e)}$ must be of the form:
\begin{equation}
    \frac{\Delta S_{s(e)}}{k_B} = c_{s(e)} + \log N,
\end{equation}
where $c_{s(e)}$ is a defect-type-dependent constant. For comparison, this constant vanishes for simple point defects in periodic crystals.

Combining these expressions with Eqs. \ref{eq:F=U-TS} and \ref{eq:DeltaU} (rewritten in terms of $\epsilon_{s(e)}$), we arrive at an estimate for the total configurational entropy of the tiling in the low-defect limit:
\begin{equation}\label{eq:entropy-ns-ne}
\begin{aligned}
    \frac{1}{k_B} S(N, n_s, n_e) =\; &\frac{1}{k_B} S_{\mathrm{sq\text{-}tr}}(N) \\
    &+ n_s \left[1 + c_s - \log\left(\frac{n_s}{N}\right) \right] \\
    &+ n_e \left[1 + c_e - \log\left(\frac{n_e}{N}\right) \right].
\end{aligned}
\end{equation}
This expression provides a baseline for understanding the entropic impact of dilute defects and offers a valuable tool for calculating equilibrium defect densities in e.g. particle-based simulations.

Another useful perspective is to analyze the configurational entropy as a function of the total number of defects, $n_D$, without distinguishing between defect types. To achieve this, we first rewrite Eq.~(\ref{eq:entropy-ns-ne}) in terms of $n_D$ and the difference between the number of shield and egg defects:
\begin{equation}
    n_s = \frac{1}{2}\left( n_D +  \Delta_D\right),\quad n_e = \frac{1}{2}\left( n_D -  \Delta_D\right),
\end{equation}
where $\Delta_D = n_s - n_e$.
We then maximize the entropy with respect to $\Delta_D$. The resulting expression for the total entropy becomes:
\begin{equation}
    \begin{aligned}
        \frac{1}{k_B}S(N,n_D) &= \frac{1}{k_B}S_\mathrm{sq\text{-}tr} \\
        &\quad + n_D \left[1 + c_D - \log\left(\frac{n_D}{N}\right)\right],
    \end{aligned}
\end{equation}
where the constant $c_D$ encapsulates the combined contribution of both defect types and is given by
\begin{equation}\label{eq:cfomcsce}
    c_D = \log\left(e^{c_s} + e^{c_e}\right).
\end{equation}
From the properties of the logarithm of a sum of exponentials, we have the bounds
\begin{equation}
    \max\{c_s, c_e\} < c_D \leq \max\{c_s, c_e\} + \log(2),
\end{equation}
with the upper bound reached only when $c_s = c_e$. This result shows that the presence of multiple defect types enhances the overall entropy, regardless of asymmetries in their numbers or individual entropic contributions. The system gains additional configurational freedom by mixing different types of defects, which further stabilizes the tiling.

\subsection{Results}

\subsubsection{Checking for quasicrystallinity and finite-size effects}

\begin{figure*}
    \centering
\includegraphics[width=\linewidth]{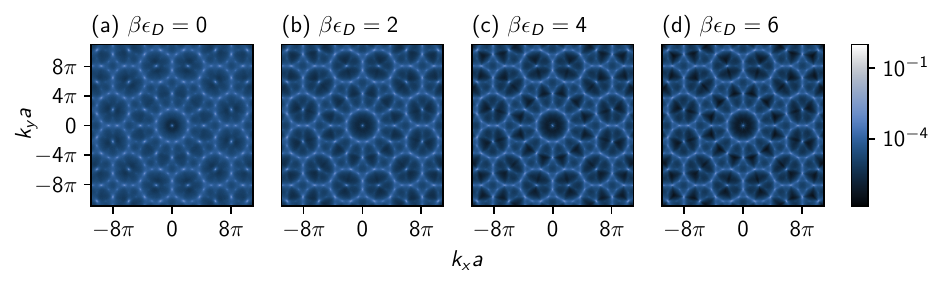}
    \caption{Ensemble averaged 2D structure factor of the simulated square-triangle-defect tilings at $14068$ vertices, $\beta\gamma a=4$, and various defect costs as indicated.}
    \label{fig:Sk}
\end{figure*}

Before conducting the configurational entropy analysis, we first confirm that the tiling remains quasicrystalline throughout the chosen simulation parameter range. To verify this, we compute the two-dimensional structure factor $S(\mathbf{k})$  of the vertices in equilibrium snapshots to examine the orientational ordering of the system. As shown in Fig.~\ref{fig:Sk}, the structure factor exhibits clear 12-fold rotational symmetry across all simulated conditions. This indicates that the tilings maintain their quasicrystalline nature throughout the simulation.

\begin{figure}
    \centering
    \includegraphics[width=\linewidth]{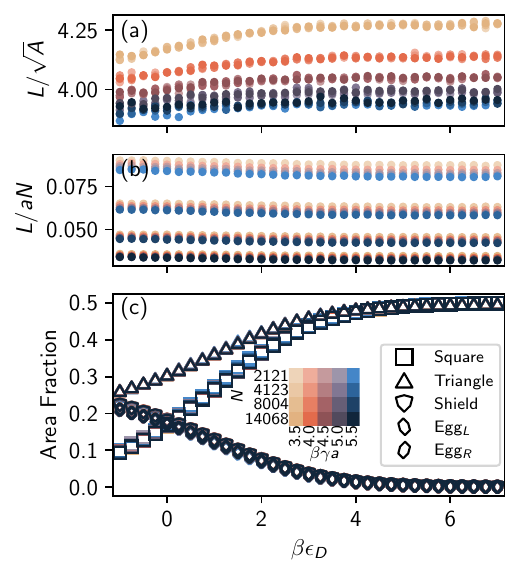}
    \caption{Boundary scaling and tile statistics in the lattice model.
(a) Ensemble-averaged perimeter scaled by the square root of the tiling area as a function of defect cost.
(b) Ensemble-averaged perimeter normalized by the number of vertices.
(c) Area fractions of squares, triangles, shields, and eggs as a function of defect cost.
Data are shown for multiple system sizes and line tensions, which are color-coded as indicated in the legend. Note that in (c) the data for all three defect types overlap.
}
    \label{fig:perimeter}
\end{figure}

In order to further confirm quasicrystallinity, we also examine the boundary behavior to ensure that the tilings remain compact across the explored range of line tensions. To assess this, we calculate the ensemble-averaged perimeter scaled by both the square root of the tiling area, $A$, and the number of vertices as a function of the defect energy penalty $\epsilon_D$, as shown in Fig.~\ref{fig:perimeter}.
In Fig.~\ref{fig:perimeter}(a), we observe that when the perimeter is scaled by the square root of the tiling area, the curves collapse for individual line tension values, regardless the system size. Additionally, the scaled perimeter decreases monotonically with increasing line tension, indicating that stronger boundary confinement suppresses elongation. In Fig.~\ref{fig:perimeter}(b), we plot the perimeter normalized by the number of vertices. This measure remains relatively constant across all defect energy penalties and decreases systematically with increasing system size and line tension.
Taken together, these observations confirm that the boundary contributions become negligible in the thermodynamic limit, ensuring that the configurational entropy analysis is minimally affected by finite-size effects.

\subsubsection{Tile concentrations}

We now explore the behavior of the concentration of
different tile types as a function of the biasing strength $\epsilon_D$
which tunes the propensity of the system to include shield and egg tiles.
Specifically, in Fig. \ref{fig:perimeter}(c), we plot the fraction of the total
area covered by squares, triangles, shields, and eggs as a function
of $\beta\epsilon_D$. We immediately observe that the area fractions are almost independent of the system size and line tension throughout the simulated ranges. Furthermore, we find that for sufficiently large $\beta\epsilon_D \gtrsim 6$, the area fraction of the defects becomes vanishingly small and the squares and triangles span equal areas of the tiling. This behavior is essentially the defect-free square-triangle tiling limit and is consistent with the geometric constraints, i.e., that the area fractions of squares and triangles in a globally uniform 12-fold symmetric square-triangle tiling must be equal \cite{oxborrow1993random,imperor2021square}.

Direct derivation of the equivalent geometric constraints, as presented in the literature \cite{oxborrow1993random,ulugol2025defects,imperor2024higher}, for square-triangle-shield-egg tilings is problematic since the \emph{representative surfaces} \cite{oxborrow1993random} of shield and egg tiles are non-linear. Thus, they break the linearity assumption of the Nienhuis relation \cite{nienhuis1998exact,imperor2021square}.
\begin{figure}
    \centering
    \includegraphics[width=0.9\linewidth]{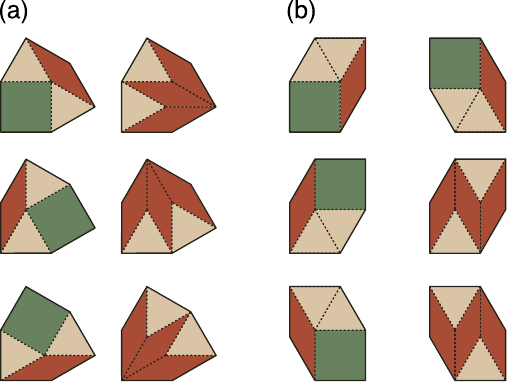}
    \caption{Complete list of square-triangle-rhombus decompositions of the defect tiles. In (a) we show 6 unique representations of the shield tile, while the two columns of (b) shows the 3 unique representations of the left- and right-handed egg respectively.
}
    \label{fig:defect-decorations}
\end{figure}
Fortunately, we can circumvent the need for a direct derivation by decorating shield and egg tiles with square, triangle, and thin rhombus tiles. The decorations of shield and egg tiles are not unique as we show in Fig.~\ref{fig:defect-decorations}. Nonetheless, we point out that the geometric constraints remain invariant of the decoration choices, and are given by:
\begin{align}
        \tau + \frac{1}{2}\left(\sqrt{3} - 1\right)\left(\Sigma + \eta_L 
    + \eta_R\right) &= \frac{1}{2}-\delta (N,\beta\epsilon_D),\label{eq:triangle-constraint}\\
    \sigma + \frac{1}{2}\left(3-\sqrt{3} \right)\left(\Sigma + \eta_L 
    + \eta_R\right) &= \frac{1}{2}+\delta (N,\beta\epsilon_D), \label{eq:quad-constraint}
\end{align}
where $\sigma$, $\tau$, $\Sigma$, $\eta_L$, $\eta_R$, represent the area fractions of square, triangle, shield, left- and right-handed egg tiles, respectively, and $\delta (N,\beta\epsilon_D)$ is a  function that encodes the deviations from the ideal case such as finite-size effects (see Appendix \ref{ap:geoconst} for a detailed discussion). Note that $\delta (N,\beta\epsilon_D)$ is expected to vanish in the thermodynamic limit. 
We term the left hand sides of  equations \ref{eq:triangle-constraint} and \ref{eq:quad-constraint} as triangle relation and quadrilateral relation, respectively.

\begin{figure}
    \centering
    \includegraphics[width=\linewidth]{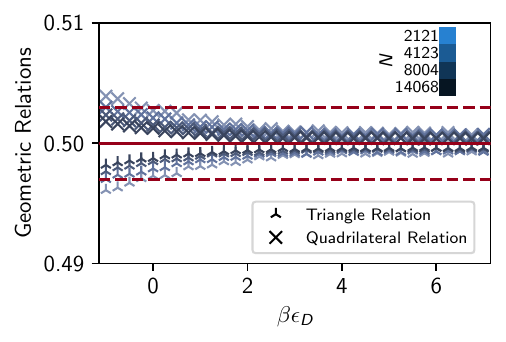}
    \caption{Line tension and ensemble averaged geometric relations as a function of defect cost. The solid horizontal line is drawn at $1/2$ to represent the ideal value which the infinite tiling must satisfy and dashed horizontal lines represent the $\pm 0.003$ deviations from the ideal value. }
    \label{fig:geometric-constraint}
\end{figure}

To check if the geometric constraints are satisfied, we plot the triangle and quadrilateral relations as a function of defect cost, $\beta\epsilon_D$, for all simulated system sizes in Fig.~\ref{fig:geometric-constraint}. Since the values are essentially independent of the line tensions, we take an average over line tensions per data point for better statistics. We find that for defect costs $\beta\epsilon_D > 0$ the triangle and quadrilateral relations remain bounded within $[0.497, 0.5]$ and $[0.5, 0.503]$, respectively. Both relations tend to $0.5$ as the system size, $N$, increases since the finite-size effects diminish. Therefore, the behavior of  the sampled tilings is consistent with what we would theoretically expect for a globally uniform 12-fold symmetric tiling.

\subsubsection{Configurational entropy of the tiling}

\begin{figure}
    \centering
    \includegraphics[width=\linewidth]{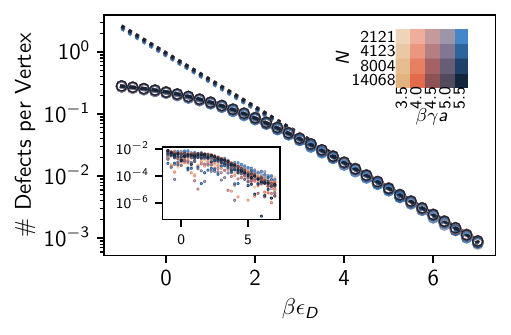}
    \caption{Total number of defects as a function of defect cost. The solid curves depict the rational function fits to the data and the dashed curves show the exponential fits to the high-defect-cost regime following Eq.~(\ref{eq:defect-scaling-high-epsilon}). The inset shows the absolute difference of the number of defects from the reference state $N=4123$ and $\beta\gamma a = 4$. The data and curves are color-coded as shown in the figure.}
    \label{fig:defect-num}
\end{figure}

A key question in this study is how the presence of shield and egg tiles affects the configurational entropy of the tiling. To quantify this effect, we compute the entropy using Eq.~(\ref{eq:F=U-TS}), Eq.~(\ref{eq:DeltaU}), and Eq.~(\ref{eq:thermo-integrate}), with Eq.~\ref{eq:sconf-def-free} serving as a reference point. However, to carry out this calculation, we require explicit functional forms for the ensemble-averaged perimeter, $\langle L\rangle$, and the total number of defect tiles, $\langle n_D\rangle$, as they appear in Eq.~(\ref{eq:DeltaU}) and Eq.~(\ref{eq:thermo-integrate}). Following the approach outlined in Ref.~\cite{ulugol2025defects}, we model $\langle L\rangle$ as a generalized logistic function of $\beta\epsilon_D$ and $\langle n_D\rangle$ as a rational function of $e^{-\beta\epsilon_D}$. These functional forms provide an excellent fit to our simulation data, as illustrated in Fig.~\ref{fig:defect-num}, where the fitted curves for the defect concentration are shown as solid lines.

\begin{figure}
    \centering
     \includegraphics[width=\linewidth]{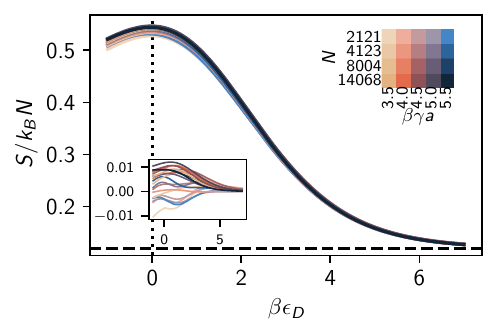}
    \caption{Configurational entropy per vertex as a function of the defect cost. The horizontal dashed line represents the baseline configurational entropy per vertex of a pure square-triangle tiling and the vertical dotted line highlights the zero defect-cost case where the configurational entropy is maximized. The inset shows the difference of the configurational entropies from the reference state $N=4123$ and $\beta\gamma a = 4$.  The curves are color-coded as shown in the legend.}
    \label{fig:s-v-eps}
\end{figure}

Using these fits, we compute the configurational entropy of the sampled tilings as a function of $\beta\epsilon_D$ and present the results in Fig.~\ref{fig:s-v-eps}. As a baseline, the configurational entropy of an infinite-size, defect-free square-triangle tiling is included as a dashed line.

We observe that, regardless of system size and $\gamma$, the configurational entropy per vertex increases significantly as the defect energy penalty $\beta\epsilon_D$ decreases. The entropy reaches its maximum when there is no defect cost ($\beta\epsilon_D = 0$), reflecting the maximally random state.

\begin{figure}
    \centering
    \includegraphics[width=\linewidth]{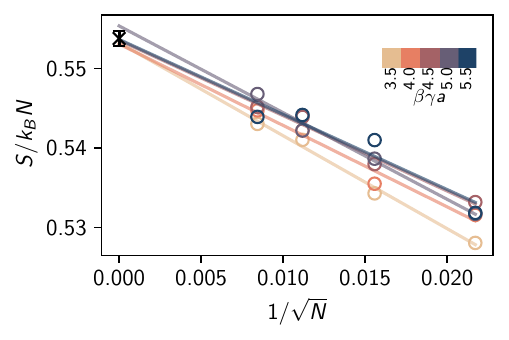}
    \caption{The finite size scaling of the configurational entropy at no defect penalty. The lines depict the linear fits to the scaling and the black $\times$ marker shows the infinite system size extrapolation value, ie, $S = 0.554(1) k_B $. The data is color-coded with respect the different line tensions as shown in the legend.}
    \label{fig:smax-finite-size}
\end{figure}

To further analyze this behavior, we perform finite-size scaling of the entropy at $\beta\epsilon_D = 0$ for all simulated line tensions, shown in Fig.~\ref{fig:smax-finite-size}. Linear fits to system sizes confirm that all systems extrapolate consistently to the infinite system size limit ($1/N = 0$). From this, we determine the configurational entropy of the infinite system to be $0.554(1)k_B$ per vertex, which is approximately $5$ times the configurational entropy of the defect-free square-triangle tiling. Here and throughout the paper, uncertainties in parentheses indicate one standard error in the last reported digit..

\begin{figure}
    \centering
    \includegraphics[width=\linewidth]{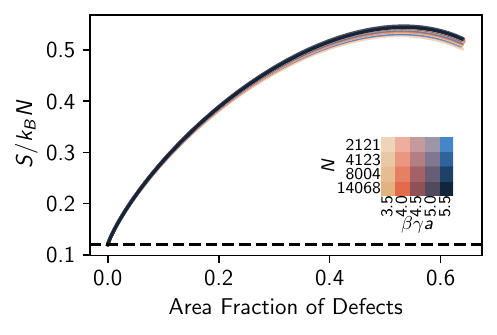}
    \caption{Configurational entropy per vertex as a function of the area fraction of defects. The horizontal dashed line represents the baseline configurational entropy per vertex of pure square-triangle tiling. The curves are color-coded as shown in the legend.}
    \label{fig:s-v-frac}
\end{figure}

While $\epsilon_D$ serves as a useful control parameter in our simulations, a more intuitive measure is the relationship between entropy increase and defect concentration. To this end, we replot the entropy data from Fig.~\ref{fig:s-v-eps} as a function of the defect area fraction, shown in Fig.~\ref{fig:s-v-frac}. A sharp increase in entropy is observed with even a small introduction of defects, reaching a maximum when the defect concentration is $0.537(1)$ corresponding to $\beta\epsilon_D = 0$.

\subsubsection{Configurational entropy of a single defect}

The sharp increase in configurational entropy at low $\beta\epsilon_D$, as seen in 
Fig.~\ref{fig:s-v-frac}, highlights the significant entropic contribution of defects in 
square-triangle tilings. To quantify this contribution, we analyze the entropy gained by adding 
a single shield or egg defect to an otherwise defect-free tiling.

\begin{figure}
    \centering
    \includegraphics[width=\linewidth]{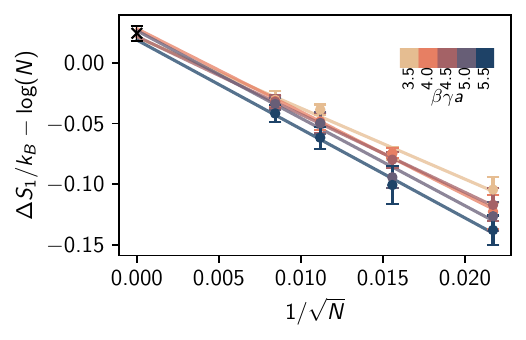}
    \caption{The finite size scaling of the configurational entropy added to the tiling by a single defect. The lines depict the linear fits to the scaling and the black $\times$ marker shows the infinite system size extrapolation value, ie, $S = 0.024(3) k_B $. The data is color-coded with respect the different line tensions as shown in the legend.}
    \label{fig:dS1-scaling}
\end{figure}

In Fig.~\ref{fig:defect-num}, we plot the equilibrium defect concentration $n_D/N$ as a 
function of $\beta\epsilon_D$ on a logarithmic scale. In the high-$\beta\epsilon_D$ regime, where 
defects are sparse and do not interact, the data follows the expected exponential scaling given by Eq.~(\ref{eq:defect-scaling-high-epsilon}), as shown by the dashed lines. This allows
us to extract the single-defect entropy contribution $\Delta S_1$. Extrapolating to the 
infinite-system-size limit, as shown in Fig.~\ref{fig:dS1-scaling}, we obtain
\begin{equation}
    \frac{\Delta S_1}{k_B} = 0.024(3) + \log (N).
\end{equation}
This result is considerably larger than the $\log N$ scaling expected for a simple vacancy in a 
crystalline system.

For context, prior studies on square-triangle tilings have shown that introducing an additional 
vertex contributes approximately $0.12k_B$ to the configurational 
entropy~\cite{widom1993bethe,kalugin1994square}. Our results indicate that introducing a single 
shield or egg defect contributes approximately $1/6$ times that amount, demonstrating that defects 
play a crucial role in increasing the number of accessible tiling configurations. The enhanced 
entropy contribution suggests that defects are not merely tolerated within the tiling but actively 
promote configurational freedom, which in turn affects the equilibrium defect concentration.

\begin{figure}
    \centering\includegraphics[width=\linewidth]{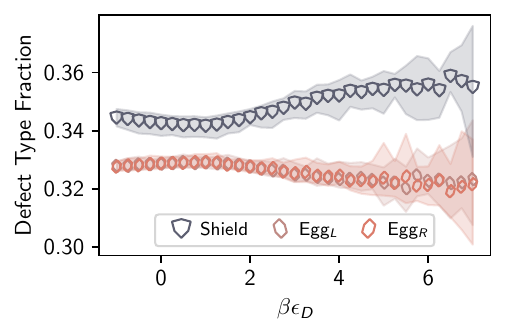}
    \caption{System size, line tension and ensemble averaged fractions of shield and egg defects among all defects as a function of defect cost.
    \label{fig:shield-egg-frac}}
\end{figure}

While the total defect concentration exhibits a consistent exponential scaling, a closer inspection, given in Fig.~\ref{fig:shield-egg-frac}, reveals that shield and egg defects are not equally favored. Thus, we separately fit the high-$\beta\epsilon_D$ regime for each defect type, extracting their individual constants $c_s$, $c_{e_L}=c_{e_R}=c_{e}-\log(2)$, and  in the entropy scaling expression $\Delta S / k_B = c + \log N$.

Intriguingly, we find that both shield and egg defects yield negative constants: $c_s=-0.97(1)$, $c_e=-0.42(1)$; suggesting that, when considered in isolation, these individual defects are actually less favorable than typical point defects in periodic crystals, which usually contribute $\log N$. 
However, when the three defect types are allowed to mix freely, the configurational entropy associated with choosing among distinct defect types introduces a significant additional entropy contribution. This mixing entropy elevates the total entropy gain to a level that not only compensates for the negative constants but results in a net favorable entropy increase per defect, as reflected in the positive $c_D=0.024(3)$ extracted from the total defect concentration.

As a consistency check, we use Eq. (\ref{eq:cfomcsce}) to estimate the constant $c_D$ from the independently measured values of $c_e$ and $c_s$. Propagating the corresponding uncertainties, we obtain $c_D = 0.035(7)$, which is consistent with the directly measured value within statistical uncertainty.


These results highlight a key insight: it is not the presence of a particular defect, but rather the freedom to mix and rearrange among multiple defect types that increases configurational entropy. 

\section{Defects in Quasicrystals Formed by Core-Corona Particles}

In the previous section we provided a method to calculate the excess configurational entropy for tilings that contain shield and egg defect tiles. In this section, we make use of this result to explore the prevalence of these defect tiles in a simple coarse-grained model of square-shoulder particles, which is well known to form a square-triangle tiling \cite{dotera2014mosaic,pattabhiraman2015stability}.

\subsection{Methods}

\subsubsection{Model}

\begin{figure}
    \centering
    \includegraphics[width=\linewidth]{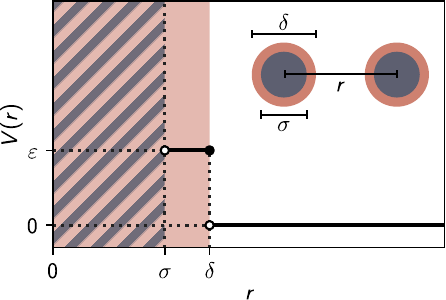}
    \caption{Schematic representation of the square-shoulder potential. The gray regime shows the hard-core of the particle while the pink regime represents the `deformable’ corona. }
    \label{fig:sqsh}
\end{figure}

We consider particles interacting via the square-shoulder pair potential, illustrated in Fig.~\ref{fig:sqsh}, defined by:
\begin{equation}\label{eq:sq_shoulder}
V(r_{ij}) = 
    \begin{cases} 
      \infty & r_{ij}\leq \sigma, \\
      \epsilon & \sigma < r_{ij} \leq \delta, \\
      0 & \delta < r_{ij},
   \end{cases}
\end{equation}
where $r_{ij} = \left|\vec{r}_j-\vec{r}_i\right|$ is the interparticle distance, $\sigma$ is the particle diameter, $\epsilon$ is the shoulder height, and $\delta = 1.4\sigma$ is the shoulder width. For this choice of $\delta$, the system exhibits 
a fluid phase, square (sq) and  hexagonal (hex) crystal phases, as well as 
a stable 12-fold symmetric quasicrystalline phase in a narrow range of densities for $k_BT\lesssim 2\epsilon$, between the region of stability of the square and hexagonal crystals \cite{pattabhiraman2015stability,pattabhiraman2017phase}.

\subsubsection{Free energy of the square and hexagonal lattices}

In order to determine the free energy of the defect-free crystal phases we use standard Einstein integration \cite{frenkel1984new} combined with thermodynamic integrations over the pressure and interaction strength.

\subsubsection{Free energy of the quasicrystal}

Calculating the free energy of a quasicrystal is more complex than that of a periodic crystal due to the many different tilings that need to be sampled. Specifically, the partition function can be written as the sum over the set $\mathcal{T}$ of distinct tile arrangements as follows:
\begin{equation}
    Z = \frac{1}{\Lambda^{3N} N!} \int \mathrm{d}\mathbf{r}^N \exp(-\beta U(\mathbf{r}^N)) = 
    \sum_{t \in \, \mathcal{T}} Z_\mathrm{vib}(t) 
\end{equation}
with 
\begin{equation}
     Z_\mathrm{vib} (t) = \frac{1}{\Lambda^{3N} N!}\int_t \mathrm{d}\mathbf{r}^N \exp(-\beta U(\mathbf{r}^N)). 
\end{equation}
Here $\mathbf{r}^N$ indicates the position of all the particles, the subscript $t$ indicates an integration over all particle configurations associated with the specific tiling $t$, $\Lambda$ is the thermal wavelength, and $U$ is the potential energy. Note that here $N$ refers to the number of particles. However, as the particles sit at the vertices of the quasicrystal tiling, the number of particles equals the number of vertices.  Additionally, $Z_\mathrm{vib}(t)$ is the vibrational free energy associated with tiling $t$: it captures all the fluctuations of particles around their equilibrium positions in that specific tiling. This can be determined using the same methods as for periodic crystals, i.e. Einstein integration \cite{frenkel1984new}. For the quasicrystal, Einstein integration yields the vibrational free energy only, which may depend on the specific realization of the tiling.

In the \textit{specific case} where the vibrational free energy is independent (or almost independent) of the tiling, the partition function can be approximated as:
\begin{equation}
    Z = Z_\mathrm{vib} Z_\mathrm{conf} \label{eq:ZZZ}
\end{equation}
where $Z_\mathrm{vib}$ is the vibrational contribution of any single tiling. The total free energy can then be obtained by summing the contribution given by the configurational entropy (we consider the per-particle value in Eq.~(\ref{eq:sconf-def-free}), obtained in the thermodynamic limit) with the vibrational free energy of a single tiling.

For the system studied here, we find that this factorization holds to within our numerical precision, justifying the assumption in Eq. \ref{eq:ZZZ}. Specifically, we computed $F_\mathrm{vib}$ for 10 different square-triangle random tilings generated with the algorithm proposed in Ref.~\onlinecite{oxborrow1993random}. We found that the free-energy differences between these tilings is small compared to our statistical uncertainty for the vibrational free energy of any given tiling (obtained by recalculating $F_\mathrm{vib}$ for the same tiling 10 times). We can then consider them as essentially identical.

To calculate the free energy at different densities, we integrated along the equation of state $P(\rho)$:
\begin{equation}\label{eq:integration-rho}
    \frac{F(\rho) - F(\rho_\text{ref})}{N k_B T} =  \int_{\rho_\text{ref}}^\rho \frac{\beta P(\rho^\prime)}{{\rho^\prime}^2}\mathrm{d\rho^\prime}
\end{equation}
where the pressure $P$ was evaluated at constant $N$, $T$, and surface area $A$ in event-driven molecular dynamics (EDMD) simulations \cite{rapaport2004art,smallenburg2022efficient}. Average pressure values and the respective statistical errors were calculated over 10 independent runs with different initial tilings. 

Using the methods outlined above, we first calculated the free energies of all relevant phases at temperature $k_B T / \epsilon = 0.1$.
To calculate the free energy at different temperatures, we again perform thermodynamic integration \cite{frenkel2023understanding}, now along linear paths in the $(\rho\sigma^2,k_BT/\epsilon)$ plane following:

\begin{equation}
    \rho(\epsilon) = \rho_1 - \frac{C}{\sigma^2}\left(\frac{k_B T}{\epsilon}- \frac{k_B T}{\epsilon_1}\right)\label{eq:integrationpath}.
\end{equation}
 
Here, $C$ was chosen to ensure that the entire integration path remained within the equilibrium region of the given phase, based on the phase diagram from Ref. \cite{pattabhiraman2017phase}, while $(\rho_1,\epsilon_1)$ is the reference state point at which we performed vibrational free energy calculations.
Tab.~\ref{tab:integ_extrema} lists the integration bounds for the three phases. The integrals for the thermodynamic integration were evaluated with Simpson's rule, with 21 points along the paths between $k_BT/\epsilon=0.1$ and $0.2$, and 11 points between $k_BT/\epsilon=0.2$ and $0.3$, including the initial and final states.

\begin{table}[]
    \centering
    \renewcommand{\arraystretch}{1.5}
    \setlength{\tabcolsep}{12pt}
    \begin{tabular}{|c||c|c|c|}
        \hline
        \hline
         $k_BT/\epsilon$ & $\rho_\text{sq}\sigma^2$ & $\rho_\text{qc}\sigma^2$ & $\rho_\text{hex}\sigma^2$ \\
        \hline
        \hline
         0.1 & 0.976 & 1.0485 & 1.122 \\
         \hline
         0.2 & 0.9395 & 1.0186 & 1.092225 \\
         \hline
         0.3 & 0.9165 & 0.9906 & 1.0823 \\
         \hline
         \hline
    \end{tabular}
    \caption{Reference particle densities used for the calculation of free energies of the three phases at the studied temperatures. The free energy values at higher temperatures were obtained by thermodynamic integration along linear paths connecting these state points.}
    \label{tab:integ_extrema}
\end{table}

\subsubsection{Free energy of single vacancies and single defects}
To calculate the free-energy cost of creating vacancies in the quasicrystalline phase, we adapt a standard procedure commonly used for periodic crystals \cite{frenkel2023understanding}. Within this framework, the free-energy cost associated with a single vacancy is given by:
\begin{equation}\label{eq:f_vac}
    -f^d = F_N(N-1,A) - F_N(N,A)
\end{equation}
where $F_M(N,A)$ denotes the free energy of a system containing $N$ particles and $M$ lattice sites, where any vacancies are placed at specific sites (i.e. the configurational entropy is not included in this free energy). 
Assuming that the pressure $P$ of the system is not strongly affected by the presence of defects, the vibrational free-energy cost of a vacancy created at constant $NPT$ is approximately \cite{frenkel2023understanding}:
\begin{equation}\label{eq:g_vac}
    g^d = G_{N+1}(N) - G_N(N) =  \mu - f^d
\end{equation}
where $G_{M}(N)$ is the Gibbs free energy for the system with $M$ lattice sites and $N$ particles again excluding any configurational entropy. Here, $\mu$ denotes the chemical potential of the defect-free lattice at the same thermodynamic conditions.

We already know $F_N(N,A)$ and $\mu$ for the defect-free case from our earlier free-energy calculations. In principle, $F_N(N-1,A)$ can be determined via a similar Einstein integration procedure \cite{frenkel1984new}. However, two complications arise for the quasicrystalline lattice, as we will discuss below. \\

\noindent \textbf{Complication 1: Vacancy defects split into pairs.}
In square-triangle quasicrystals, a vacancy does not remain localized at a single lattice site. As  illustrated in Fig.~\ref{fig:defectdyn}, it relaxes into a pair of independent structural defects (i.e. eggs or shields), through local rearrangements that preserve the 12-fold orientational order. Consequently, in our simulations, the quantity $F_N(N-1,A)$ needs to be computed for configurations containing such \textit{pairs} of defective tiles.

When the concentration of defective tiles is low and the average separation is sufficiently large, it is reasonable to assume that they do not interact and contribute independently to the free energy. Under this assumption, the vacancy free energy can be written as:
\begin{equation}\label{eq:g_defects}
    f^d = f^\text{e/s}_1 + f^\text{e/s}_2 \quad.
\end{equation}
where $f^\text{e/s}$ is the free-energy cost associated with an individual egg or shield defect, and the subscript acts as a label for each individual defect.

\begin{figure}
    \centering
    \includegraphics[width=0.8\linewidth]{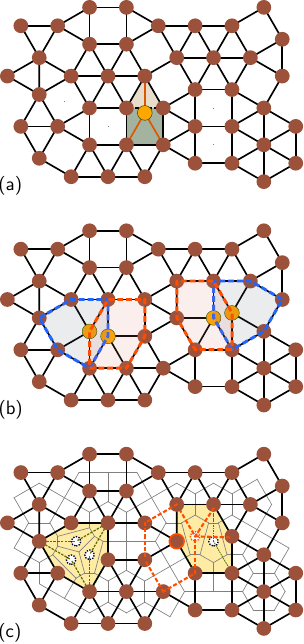}
    \caption{(a) First step of the algorithm of Ref.~\cite{oxborrow1993random}, involving the insertion of an extra vertex that creates a pair of rhombus tiles, which are subsequently propagated across the system through specific tile-flipping rules. (b) The removal of one of the two vertices at the obtuse angle of each rhombus generates either an egg (highlighted by dashed orange lines) or a shield (highlighted by dashed blue lines), each constituting half a vacancy. (c) The inner sites, illustrated as white circles with dashed contours, represent all virtual sites that would decompose defect tiles into a rhombus, a square, and two triangles, as in Fig.~\ref{fig:defect-decorations}. The orange dashed lines show the tile rearrangement obtained when a particle sitting at the obtuse angle of a defect (outlined in orange) hops to its nearest empty site, generating a different configuration. Thin solid lines represent the cells each particle is restricted to when performing simulations where we prevent defect diffusion.}
    \label{fig:defects-creation}
\end{figure}

Initial configurations containing a pair of defects were generated using a variation of the algorithm of Ref.~\cite{oxborrow1993random}. An excess vertex is first introduced into the system, replacing a square and a triangle with a pair of rhombi and a triangle (Fig.~\ref{fig:defects-creation}(a)).
The rhombi are then propagated throughout the system according to a set of local rules that preserve the 12-fold symmetry and ensure proper sampling of the configurational space.
We halt the propagation process once the rhombi are separated by a distance approximately equal to half of the linear size of the system. Subsequently, one vertex at an obtuse angle is removed from each rhombus yielding a configuration with $N-1$ occupied sites and two defects which each can either be an egg or a shield, as shown in Fig.~\ref{fig:defects-creation}(b). Initial configurations with different combinations of eggs and shields were selected. \\
  
\noindent\textbf{Complication 2: Defects diffuse through the system.}
An important issue for the defected configurations is the fact that we want to avoid sampling multiple realizations of the tiling, as we are purely interested in the vibrational free energy at this stage. In other words, we want to ensure that the defects do not diffuse during the free-energy calculations. For a crystal, this could be accomplished by restricting the movement of each particle to its Voronoi cell as calculated for a perfect defect-free lattice \cite{van2017diffusion, van2020high}. Here, we construct an analogous approach for the quasicrystalline lattice. We start by observing that each defect tile can be decomposed into two triangles, a square and a rhombus.  For the shield tile, there are three ways to do this while for each egg there are two ways (see Fig. \ref{fig:defect-decorations}), with each decomposition resulting in one additional ``virtual'' lattice site inside the defect. When a defect diffuses through the lattice, it does so via hopping of a neighboring particle onto one of these virtual sites. Hence, in order to restrict each particle to its current lattice position, we restrict its motion such that it can go no closer to any other lattice site (including virtual sites) than to its own ideal lattice site. This is illustrated in Fig.~\ref{fig:defects-creation}(c).

\subsubsection{Obtaining defect free energies at different temperatures and densities}

In practice, the method described above works best when the fluctuations in the quasicrystal structure are small enough to ensure that the restriction of particles to their local cell does not noticeably affect the free energy of the quasicrystal. This assumption is most accurate at low temperatures, where the quasicrystal is rigid. Specifically, we have confirmed that this restriction does not measurably affect the free energy of a defect-free quasicrystal at $k_B T/\epsilon = 0.1$.

Hence, in practice we calculate our defect free energies at the state point $(\rho^\text{lat}\sigma^2,k_BT/\epsilon) = (1.0485, 0.1)$ (inside the stable quasicrystal regime \cite{pattabhiraman2015stability}). 
In order to calculate vacancy free energies at higher temperatures $k_BT/\epsilon=0.2$ and $0.3$, we again employed thermodynamic integration along the same integration path of Eq.~(\ref{eq:integrationpath}). Note that additional small compressions and expansions were included to correctly incorporate the effect of the presence of vacancies on the lattice site density. 

Additionally, in order to calculate defect free energies at different densities, we first notice that within the quasicrystal region, the density $\rho$ only varies within a narrow range (the relative difference between the two densities at the two phase boundaries is between 0.1\% and 1\% for all three considered temperatures). Conversely, the pressure $P$ shows significant changes in this density range. This affects the two terms in Eq.~(\ref{eq:g_vac}) differently. The chemical potential $\mu$ for the defect-free lattice undergoes a significant variation, which to first order is proportional to $P$. The free-energy cost $f^d$ changes much more slowly, since it is only sensitive to changes in the lattice spacing, which remains nearly constant. We can then write:
\begin{equation}\label{eq:g-vac-phase-boundaries}
    g^d(\rho) = \mu(\rho) - f^d(\rho_\text{ref}) + \mathcal{O}(|\rho-\rho_\text{ref}|)
\end{equation}
where $\mu(\rho)$ was already known from our defect-free free-energy calculations. We verified the validity of this approximation in one case, $k_BT/\epsilon=0.1$, confirming that the variation of $\mu$ is  around $\sim1k_BT$, while the corresponding variation of $f^d$ is much smaller (order of $\sim0.1k_BT$).

\subsubsection{Equilibrium defect concentration}\label{sec:defect-concentration}
To finally calculate the equilibrium concentration of defects in the quasicrystal phase, we combine the configurational entropy contribution of the defects (Eq.~(\ref{eq:entropy-ns-ne})) with the vibrational contributions obtained as described above. This yields the total Gibbs free energy $G(n_s,n_e)$ for a system with $n_s$ shields and $n_e$ eggs:
\begin{equation}\label{eq:G_def_tot}
    \beta G(n_s,n_e) = \beta G(0,0) + \beta g^e n_e + \beta g^s n_s - \frac{\Delta S(n_s,n_e)}{k_B},
\end{equation}
with $g^{e/s} = \mu - f^{e/s}$. 
By minimizing this expression with respect to $n_s$ and $n_e$, we again obtain the expressions in Eq.~(\ref{eq:defect-scaling-high-epsilon}) for the equilibrium number of defects:
\begin{equation}\label{eq:defects-avg-n}
    \langle n_s \rangle = Ne^{-\beta g^s + c_s}, \quad 
    \langle n_e \rangle = Ne^{-\beta g^e + c_e}.
\end{equation}
This leads to the following expression for the equilibrium Gibbs free energy in the presence of defects:
\begin{equation}
    \beta G(n_s,n_e) \approx \beta G(0,0) - \langle n_s\rangle -\langle n_e\rangle
\end{equation}
Following the same approach derived for crystals in Ref.~\cite{frenkel2023understanding}, we can then evaluate the effect of defects on the chemical potential and pressure at coexistence through the relations:
\begin{eqnarray}
  \beta\delta P_\text{coex}  &\approx& -\frac{\langle n_s\rangle +\langle n_e\rangle}{N(a_\text{xtl}-a_\text{qc})} \label{eq:delta-p} \\
  \beta\delta\mu &\approx& \frac{\beta\delta P_\text{coex}}{\rho_\text{xtl}} \label{eq:delta-mu}
\end{eqnarray}
where $\rho_\alpha$ and $a_\alpha=\rho_\alpha^{-1}$ are the density and the area per particle of the crystal and defect-free quasicrystal at coexistence.
Here we ignore the effects of any defects in the periodic crystal phases.

\subsubsection{Average defect-type frequencies in the low-concentration limit} \label{sec:relativedefectmethods}
As a consistency check, we also calculate the relative frequencies of shields and eggs in the limit of low concentration, by performing Monte Carlo simulations of a system with a single missing particle in the $NVT$ ensemble. To facilitate fast diffusion of the two resulting defects through the system, we include additional Monte Carlo moves that allow particles to jump directly to a neighboring lattice site. Specifically, the particle displacement is chosen as
\begin{equation}\label{eq:12-fold-set}
    \Delta\vec{r} = 2a\sin\left(\pi/12\right)
    \left(
    \begin{matrix}
        \cos\left(\theta+\frac{\pi}{12}\right) \\
        \sin\left(\theta+\frac{\pi}{12}\right)
    \end{matrix}
    \right)
    \quad , \quad \theta = \frac{k\pi}{6}
\end{equation}
where $a$ is the lattice spacing and $k \in [0,11]$ is a randomly chosen integer.

We calculate the frequencies of shields and eggs over 2000 independent configurations averaged over 10 simulation runs per state point. To determine the number of eggs and shields at each considered timestep, we first construct a list of bonds based on the average nearest neighbor distance, from which we subsequently construct all the tiles, which are classified based on their number of vertices and the sequence of their internal angles. Note that tiles that do not correspond to a square, triangle, shield, or egg are classified as ``other''.

\subsection{Results}

\begin{figure*}
    \centering
    \includegraphics[width=\linewidth]{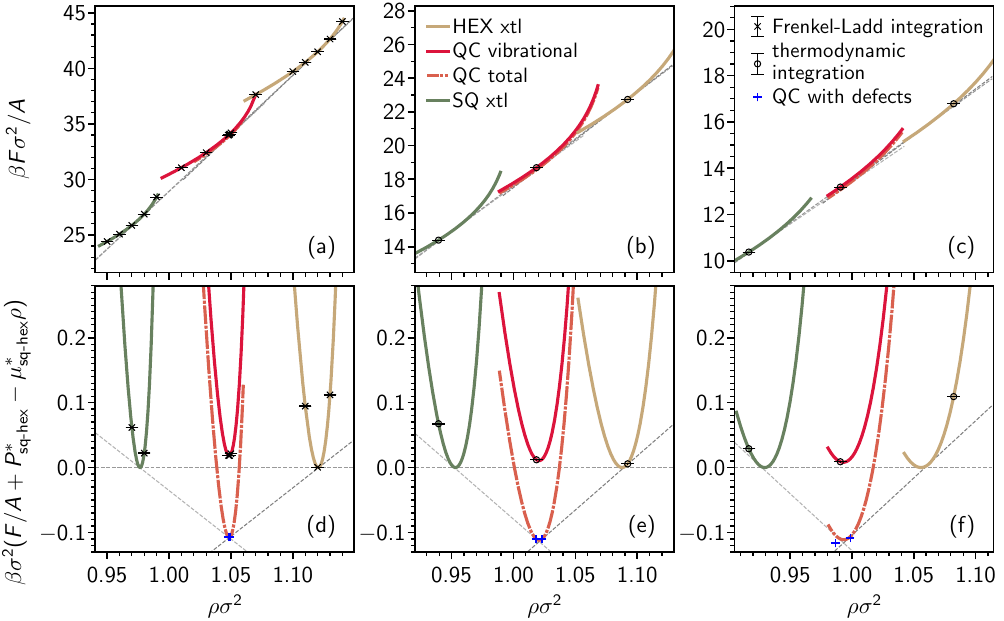}
    \caption{\textbf{Top panels}: relative free energy density $\beta F/A$ as a function of the particle density $\rho$ for square crystal (sq), 12-fold quasicrystal (qc, with and without configurational entropy contribution), hexagonal (hex) phases.
    \textbf{Bottom panels}: same data as in the top panels, but with a constant linear function subtracted corresponding to the free energy of a hypothetical coexistence between  the square and hexagonal phases at the same density. These plots highlight that the quasicrystal phase is consistently stabilized relative to the crystalline phases by its configurational entropy.
    Data are shown at three different temperatures: $k_BT/\epsilon = 0.1$ (a,d), $0.2$ (b,e) and $0.3$ (d,f). Symbols indicate free energies computed using the Frenkel–Ladd method ($\times$) at their reference densities for $k_BT/\epsilon = 0.1$, and via additional thermodynamic integration over $\rho$ and $T$ ($\circ$) at higher temperatures. Solid lines represent free energy branches obtained by thermodynamic integration over density, while the orange dash-dotted line also incorporates the configurational free energy for the quasicrystal. Dashed gray lines indicate the common tangents between phase branches, connecting coexisting states. Blue plus symbols ($+$) indicate the free energy values at coexistence taking into account the presence of defects.}
    \label{fig:fen-vs-rho}
\end{figure*}
\subsubsection{Stability region of the quasicrystalline phase}
Using the methodology described in the previous section, here we assess the full thermodynamic effect of defects on the quasicrystalline phase in a system of core-corona particles mutually interacting through the pair-potential of Eq.~(\ref{eq:sq_shoulder}). We begin by using free-energy calculations to obtain the equilibrium phase boundaries of the quasicrystal phase at temperatures $k_B T/\epsilon = 0.1, 0.2$, and $0.3$. Specifically, we  calculated the free energy density $\beta F/A$ as a function of the density $\rho$ for the square crystal, quasicrystal, and hexagonal crystal. The results are shown in Fig.~\ref{fig:fen-vs-rho}(a-c), where the dashed common-tangent lines represent the free energies of phase coexistences.
Additionally, in Fig.~\ref{fig:fen-vs-rho}(d-f) we show the same data, but with a constant slope (equal to the chemical potential of a hypothetical square-hexagonal coexistence) subtracted, in order to show the relative free energies more clearly. In particular, these figures show both the pure vibrational free-energy of the quasicrystal (solid red lines) and its total free energy (dash-dotted orange lines). Clearly, without the configurational entropy, the quasicrystal phase is not stable with respect to a square-hexagonal coexistence. Interestingly, this is in contrast with earlier calculations in Ref. \cite{pattabhiraman2015stability}. However, once the configurational entropy is included, a region emerges where the quasicrystal is thermodynamically stable. Note that a similar scenario, where configurational entropy was required for the stability of a 12-fold symmetric quasicrystal phase, was observed in a binary mixture of hard spheres on a flat substrate \cite{fayen2024quasicrystal}.

For completeness, we report the calculated transition densities and pressures at the three temperatures analyzed in Tab.~\ref{tab:qc_boundaries}, together with their statistical uncertainties.

\begin{table*}[]
    \centering
    \renewcommand{\arraystretch}{1.5}
    \setlength{\tabcolsep}{5pt}
    \begin{tabular}{|c||c|c|c|c||c|c|c|c|}
        \hline
        \hline
         $k_BT/\epsilon$ & $\rho_\text{sq-qc}\sigma^2$ & $\beta P_\text{sq-qc}\sigma^{-2}$  & $\langle n_s\rangle/N$ & $\langle n_e\rangle/N$ & $\rho_\text{qc-hex}\sigma^2$ & $\beta P_\text{qc-hex}\sigma^{-2}$ & $\langle n_s\rangle/N$ & $\langle n_e\rangle/N$ \\
        \hline
        \hline
         0.1 & 1.048022(6) & 74.203(7) & $2(1)\cdot10^{-8}$ & $5(3)\cdot10^{-8}$ & 1.049167(5) & 77.336(3) & $0.5(2)\cdot10^{-8}$ & $1.2(6)\cdot10^{-8}$ \\
         \hline
         0.2 & 1.01829(2) & 35.862(5) & $1.4(7)\cdot10^{-4}$ & $4(2)\cdot10^{-4}$ & 1.02319(2) & 39.259(4) & $3(1)\cdot10^{-5}$ & $7(4)\cdot10^{-5}$ \\
         \hline
         0.3 & 0.98742(5) & 22.997(7) & $3(2)\cdot10^{-3}$ & $9(4)\cdot10^{-3}$ & 0.99854(4) & 26.486(3) & $0.5(3)\cdot10^{-3}$ & $1.5(7)\cdot10^{-3}$ \\
         \hline
         \hline
    \end{tabular}
    \caption{Quasicrystal phase boundaries calculated at three different temperatures; columns on the left side report the coexistence density and pressure with the square crystal, together with the estimated equilibrium fraction of defective shield- and egg-shaped tiles, while the last four columns report the same quantities at the state point coexisting with the hexagonal crystal.}
    \label{tab:qc_boundaries}
\end{table*}

\subsubsection{Vibrational free-energy cost of a single defect}
Analogous to what we did in the first part of the manuscript, where we calculated the excess configurational entropy provided by the presence of defective tiles, here we calculate the vibrational free-energy cost to create a single defect in the ideal tiling at a certain pressure $P$ and temperature $T$.

\begin{table*}[]
    \centering
    \begin{subtable}[]{0.28\linewidth}
    \renewcommand{\arraystretch}{1.5}
    \setlength{\tabcolsep}{5pt}
    \begin{tabular}{|c|c|}
        \hline
        \hline
         vacancy type & $\beta F^d/N$ \\
        \hline
        \hline
        \marker{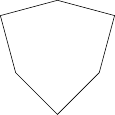} + \marker{mshield.pdf} & 32.44386(4) \\
         \hline
         \marker{mshield.pdf} + \marker{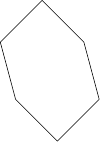} & 32.44375(5) \\
         \hline
         \marker{mel.pdf} + \marker{mel.pdf} & 32.44360(5) \\
         \hline
         \marker{mel.pdf} + \marker{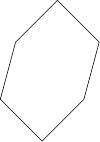} & 32.44361(5) \\
         \hline
         \hline
    \end{tabular}
    \caption{\label{tab:fen_vac_pairs}}
    \end{subtable}
    \hfill
    \begin{subtable}[]{0.3\linewidth}
    \centering
    \renewcommand{\arraystretch}{1.5}
    \setlength{\tabcolsep}{5pt}
    \begin{tabular}{|c|c||c|c|}
        \hline
        \hline
        \multirow{2}{*}{$\frac{k_BT}{\epsilon}$} & \multirow{2}{*}{$\rho\sigma^2$} 
        & \marker{mshield.pdf} & \marker{mel.pdf} \marker{mer.pdf} \\
        & & $\beta g^\text{s}$ & $\beta g^\text{e}$ \\
        \hline
         \hline
         0.1 & 1.0485 & 17.35(9) & 16.98(9) \\
         \hline
         0.2 & 1.0186 & 8.1(2) & 7.6(2) \\
         \hline
         0.3 & 0.9906 & 5.4(2) & 4.9(2) \\
         \hline
         \hline
    \end{tabular}
    \caption{\label{tab:g-vac-refpoints}}
    \end{subtable}
    \hfill
\begin{subtable}[]{0.40\linewidth}
    \centering
    \renewcommand{\arraystretch}{1.5}
    \setlength{\tabcolsep}{3pt}
    \begin{tabular}{|c||c|c||c|c|}
        \hline
        \hline
        \multirow{2}{*}{$\frac{k_BT}{\epsilon}$} 
        & \multicolumn{2}{c||}{$\beta g^\text{e/s}(\rho_\text{sq-qc})$}
        & \multicolumn{2}{c|}{$\beta g^\text{e/s}(\rho_\text{qc-hex})$} \\
        \cline{2-5}
        & \marker{mshield.pdf} & \marker{mel.pdf} \marker{mer.pdf} & \marker{mshield.pdf} & \marker{mel.pdf} \marker{mer.pdf} \\
        \hline
         \hline
         0.1 & 16.7(2) & 16.3(2) & 18.2(2) & 17.8(2) \\
         \hline
         0.2 & 7.9(3) & 7.4(3) & 9.6(3) & 9.1(3) \\
         \hline
         0.3 & 4.8(3) & 4.3(3) & 6.6(3) & 6.1(3) \\
         \hline
         \hline
    \end{tabular}
    \caption{\label{tab:g-vac-boundaries}}
\end{subtable}
\caption{Vibrational free energies of defects.
(a) Per-particle Helmholtz free energies of quasicrystal configurations with a single pair of defects at $k_BT/\epsilon=0.1$ and $\rho\sigma^2=1.0485$; each of the four rows represents a possible configuration of a vacancy, consisting of two defective tiles; data have been averaged over 10 independent simulation runs with different initial lattices.
(b) Gibbs free-energy cost for creating a defect in the quasicrystal at fixed pressure and temperature at three different reference state points within the equilibrium region for the quasicrystal.
(c) Extrapolated defect free-energy costs at the transition points with the square and hexagonal crystals.}
\end{table*}
In Tab.\ref{tab:fen_vac_pairs} we report the per-particle Helmoltz free energy of a system with a single vacancy for 4 different cases, in which we consider different combinations of shields and eggs: shield-shield, shield-egg, two eggs with the same chirality, and two eggs with different chirality. Upon closer inspection, these results are consistent with the assumption that our system is large enough (and the defects sufficiently far apart) to ensure that the two defects do not significantly interact. Specifically, we observe that in cases with two egg defects, the handedness of the eggs has no effect on the free energy of the system. Moreover, the combination of a shield and an egg results in a free energy that is almost exactly halfway between that of the shield-shield and egg-egg cases. This validates Eq.~(\ref{eq:g_defects}), in which we split the vacancy free energies into independent contributions from the single defective tiles.

Subsequently, using Eqs.~(\ref{eq:f_vac}-\ref{eq:g_defects}), we calculated the vibrational free-energy cost of shields and eggs at the three investigated temperatures, whose values are reported in Tab.~\ref{tab:g-vac-refpoints}. For all three temperatures investigated, the free-energy cost of creating a shield is about $\sim0.5k_BT$ larger than that of creating an egg.

Finally, by using Eq.~(\ref{eq:g-vac-phase-boundaries}) we calculate approximated values for $\beta g^d$ at the transition with the two crystals, which we will use in the next section to calculate the average frequency of defects at the phase boundaries of the quasicrystal. The free energy costs are summarized in Tab.~\ref{tab:g-vac-boundaries}.

\subsubsection{Equilibrium defect concentration}
Finally, we combine the calculated vibrational cost of defects in the square-should system with the configurational entropy extracted from the lattice-based simulations, to determine the expected defect density at the phase boundaries.
Using Eq.~(\ref{eq:defects-avg-n}), we compute the expected number of shields and eggs per particle, at the state points where the quasicrystal phase coexists with the two other crystalline phases.
As shown in the results summarized in Tab.~\ref{tab:qc_boundaries}, the defect frequencies exhibit a strong dependence on both temperature and density (or equivalently, pressure). The effects of density are readily visible by looking at the difference between the defect concentrations at the (low-density) square-quasicrystal phase boundary and at the (high-density) quasicrystal-hexagonal phase boundary. At the higher-density points, the defect concentration is significantly lower, which can be understood from the fact that introducing defects increases the total area of the lattice, which is more difficult at higher densities (or pressures).

Looking at the effects of temperature, we see that at low $T$, the high vibrational free-energy cost of forming defects significantly suppresses their presence in the system. As the temperature increases, $\beta g^d$ decreases and defects become progressively more favorable. At the highest investigated temperature, $k_BT/\epsilon = 0.3$, we estimate the total defect frequency per lattice site to be approximately $(\langle n_s\rangle + \langle n_e\rangle)/N \sim 0.01$. This can be compared, for example, to the vacancy concentration in hard-sphere crystals \cite{pronk2001point}, which is on the order of $10^{-4}$. Clearly, defects in this quasicrystal phase can occur at anomalously high frequencies.

Finally, we estimate the effect of these defects on the phase boundaries. This is most relevant at high temperature $k_B T / \epsilon = 0.3$ and at the square-quasicrystal phase boundary, where this effect will be strongest. Using Eqs.~\ref{eq:delta-p}–\ref{eq:delta-mu}, we predict a shift in the coexistence chemical potential, of $\beta \delta\mu \sim -0.2$, and pressure, of $\beta \delta P\sigma^{2} \sim -0.2$. These shifts are significant in comparison to the numerical uncertainties in the defect-free values of these quantities. Despite this, the equilibrium density range of the quasicrystal phase is only weakly affected, broadening by roughly $7\%$ due to the steep slope of the equation of state. Plus symbols in Fig.~\ref{fig:fen-vs-rho} visually show the free-energy and coexistence density shifts due to the presence of defects, which appear to be relevant only at $k_B T / \epsilon = 0.3$. We expect these effects to become more pronounced at even higher temperatures (as long as the quasicrystal remains stable).

\subsubsection{Relative defect frequencies}
As a consistency check, we turn our attention to the relative concentration of shield and egg defects. This relative concentration follows from two contributions: the difference in vibrational free energy for the two defect types (here calculated specifically for the square-shoulder model) and the difference in their effect on the configurational entropy (as calculated from the lattice model). If our calculations are correct and the assumption of non-interacting defects is valid, we would expect that by combining these two contributions, we obtain a reliable prediction for the relative concentration of the two defect types. This prediction can then be confirmed directly in simulations of the square-shoulder model in the $NVT$ ensemble with a single missing particle, as discussed in the Sec. \ref{sec:relativedefectmethods}. In Tab.\ref{tab:def-freqs}, we report both the measurements from these simulations and our prediction based on the combined free-energy methods. At all three investigated state points, we find excellent agreement between the two methods, validating our methodology.

\begin{table}[]
    \centering
    \renewcommand{\arraystretch}{1.5}
    \setlength{\tabcolsep}{2pt}
    \begin{tabular}{|c|c||c|c|c|c|c|c|}
        \hline
        \hline
        \multirow{2}{*}{$\frac{k_BT}{\epsilon}$} & \multirow{2}{*}{$\rho\sigma^2$} 
        & \multicolumn{2}{c|}{\marker{mshield.pdf}} & \multicolumn{2}{c|}{\marker{mel.pdf} \marker{mer.pdf}} & other\\
        \cline{3-7}
        & & Simulation & Analytic & Simulation & Analytic & Simulation \\
        \hline
         \hline
         0.1 & 1.0485 & 0.28(1) & 0.28(1) & 0.72(1) & 0.72(1) & - \\
         \hline
         0.2 & 1.0186 & 0.27(1) & 0.27(1) & 0.73(1) & 0.73(1) & 0.0003(3) \\
         \hline
         0.3 & 0.9906 & 0.27(1) & 0.26(1) & 0.72(1) & 0.74(1) & 0.012(1) \\
         \hline
         \hline
    \end{tabular}
    \caption{Average relative frequency of different types of defect tiles as measured or calculated in the limit of low defect concentration. Here, ``other'' refers to defect tiles that do not fit any of the expected shapes. These are short-lived defects arising from thermal fluctuations.}
    \label{tab:def-freqs}
\end{table}

\section{Conclusions and Perspectives}
We studied the thermodynamic effect of point defects in 12-fold quasicrystalline square-triangle tilings arising due to missing particles.  We find that removing a single particle results in the addition of two extra tile types to the square-triangle tiling: shields and eggs. Interestingly these eggs are chiral and appear in two conformations. Using thermodynamic integration on a lattice model, we determined the configurational entropy gain associated with these defects, which we find to be considerable when compared to the entropy per site in the ideal quasicrystal lattice. The ability of a single missing particle to give rise to two defects that can independently diffuse through the system and change their shape greatly enhances the configurational entropy gain due to vacancies.  Interestingly, we also find that the three defect types (shield, left-handed egg, right-handed egg) are  roughly equally present in the ideal system, with a slight preference for shields. 

In order to see how such defects affect the equilibrium thermodynamics of a simple, quasicrystal-forming model system, we determined the free-energy cost of creating defects in a square-shoulder system. By combining these calculations with our configurational entropies we calculated the average equilibrium fraction of defects per particle, and assessed their effect on the coexistence chemical potentials and pressures.  We found that the concentration of defects could be significantly higher than in a typical crystal -- reaching up to about 1\%.  We were also able to check that our free energy calculations were accurate by independently verifying that they correctly predicted the relative ratio of shield and egg defect tiles.

The methods we have used here provide an effective route for the calculation of equilibrium defect concentrations in quasicrystalline tilings. However, we emphasize that care should be taken when extending this approach to other model systems. In more complex systems, it is possible that the separation of the vibrational and configurational contributions to the free energy (Eq. \ref{eq:ZZZ}) is no longer valid. Additionally, sources of configuration entropy beyond those considered here, such as dislocations or tiles that can be decorated with different arrangements of particles, may need to be considered.

Overall, our results suggest that the high prevalence of shield- and egg-shaped defects in many self-assembled quasicrystals arises because of equilibrium thermodynamics rather than kinetic trapping. Since the diffusion of defects also plays a profound role in the dynamics of these exotic phases, this implies that accounting for the behavior of defects is a vital component in the correct description of soft matter quasicrystals.

\begin{acknowledgments}
We thank Rinske Alkemade for proofreading the manuscript. AU further thanks Randall Kamien and Stefanie D. Pritzl for fruitful discussions. AU and LF acknowledge funding from the Dutch Research Council (NWO) under the grant number OCENW.GROOT.2019.071. 
\end{acknowledgments}

\section*{Data Availability}
The data and the code that support the findings of the tiling study are openly available on Zenodo at Ref.~\cite{ulugol2026tiling} and GitHub at Ref.~\cite{ulugol2026open}

\appendix

\section{Derivation of Geometric Constraints}
\label{ap:geoconst}

Following Refs.~\cite{ulugol2025defects,imperor2024higher}, the geometric constraints of a globally uniform 12-fold symmetric square-triangle-rhombus tiling require that triangles cover exactly half of the total area, while the remaining half is covered by quadrilaterals, i.e., squares and rhombi. In terms of area fractions, this is expressed as
\begin{equation}
    \tau = \frac{1}{2},\quad \sigma + \rho = \frac{1}{2},
    \label{eq:rhomb-area-constraints}
\end{equation}
where $\tau$, $\sigma$, and $\rho$ denote the area fractions of triangles, squares, and rhombi, respectively.

To map these constraints onto the square-triangle-shield-egg tiling, we decompose the shield and egg tiles into a combination of squares, triangles, and rhombi. As shown in Fig.~\ref{fig:defect-decorations}, both shield and egg tiles can be decorated using a total of two triangles along with various combinations of squares and rhombi. Importantly, all shield and egg variants contain two triangles regardless of the specific decomposition.

Let $A_D$ denote the area of a shield or egg tile (they are equal), and $A_T$ the area of a triangle. Since each defect tile contains two triangles, the fraction of the defect tile area that maps to triangles is $2A_T / A_D$. Substituting in the tile areas $A_T/a^2 = \sqrt{3}/4$ and $A_D/a^2 = 3/2 + \sqrt{3}/2$, we obtain $2A_T / A_D = (\sqrt{3} - 1)/2$. Therefore, the fraction of the total area covered by triangles, when the defects are decomposed, maps 
\begin{equation}
    \tau \rightarrow \tau + \frac{1}{2}(\sqrt{3} - 1)(\Sigma + \eta_L + \eta_R).
\end{equation}
The remaining portion of the defect area maps to quadrilaterals (squares and/or rhombi), which gives
\begin{equation}
    \sigma + \rho \rightarrow \sigma + \frac{1}{2}(3 - \sqrt{3})(\Sigma + \eta_L + \eta_R).
\end{equation}
Substituting these mappings into Eq.~(\ref{eq:rhomb-area-constraints}), we obtain:
\begin{align}
    \tau + \frac{1}{2}(\sqrt{3} - 1)(\Sigma + \eta_L + \eta_R) &= \frac{1}{2}, \\
    \sigma + \frac{1}{2}(3 - \sqrt{3})(\Sigma + \eta_L + \eta_R) &= \frac{1}{2}.
\end{align}
These equations represent the idealized constraints in the infinite-system-size limit. In practice, however, small deviations are observed due to finite-size effects and fluctuations at the boundaries. To account for these, we introduce a term $\delta(N, \beta\epsilon_D)$ that shifts area between the two equations while preserving their sum. The final expressions read:
\begin{align}
    \tau + \frac{1}{2}(\sqrt{3} - 1)(\Sigma + \eta_L + \eta_R) &= \frac{1}{2} - \delta(N, \beta\epsilon_D), \\
    \sigma + \frac{1}{2}(3 - \sqrt{3})(\Sigma + \eta_L + \eta_R) &= \frac{1}{2} + \delta(N, \beta\epsilon_D),
\end{align}
which correspond to Eqs.~\ref{eq:triangle-constraint} and~\ref{eq:quad-constraint} in the main text.

Note that the sum of the left-hand sides of these two equations equals $\tau + \sigma + \rho + (\Sigma + \eta_L + \eta_R)$, which by construction must sum to unity. The term $\delta(N, \beta\epsilon_D)$ thus captures any residual deviation from perfect area partitioning and vanishes in the thermodynamic limit.

\bibliography{refs}

\end{document}